# Chorex: Restartable, Language-Integrated Choreographies

Ashton Wiersdorf[a]  and Ben Greenman[a]

a    University of Utah, Salt Lake City, UT, USA

**Abstract**    We built Chorex, a language that brings choreographic programming to Elixir as a path toward robust distributed applications. Chorex is unique among choreographic languages because it tolerates failure among actors: when an actor crashes, Chorex spawns a new process, restores state using a checkpoint, and updates the network configuration for all actors. Chorex also proves that full-featured choreographies can be implemented via metaprogramming, and that doing so achieves tight integration with the host language. For example, mismatches between choreography requirements and an actor implementation are reported statically and in terms of source code rather than macro-expanded code. This paper illustrates Chorex on several examples, ranging from a higher-order bookseller to a secure remote password protocol, details its implementation, and measures the overhead of checkpointing. We conjecture that Chorex's projection strategy, which outputs sets of stateless functions, is a viable approach for other languages to support restartable actors.

**ACM CCS 2012**

- **Computing methodologies** → **Concurrent programming languages**;
- **Software and its engineering** → **Extensible languages**; *Compilers*;

# The Art, Science, and Engineering of Programming

## 1 Introduction

Choreographic programming adds a layer of organization to concurrent or distributed systems. A choreographic language introduces a domain-specific notation for *choreographies*—programs that describe the interactions among actors in a system—and it *projects* each choreography to a set of local programs, one for each actor [7, 39, 40]. These local programs contain all the behavior to ensure their corresponding actors fulfill their part in the choreography.

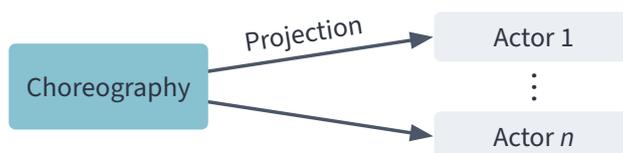

A choreography makes a global view of the system explicit as code, and is deeply connected to the actual behavior of actors. By contrast, in traditional distributed systems, the global view is merely a design document or a sketch on a whiteboard, and it is up to programmers to ensure that individual actors work together to realize the global protocol design. Actors can easily fall out of sync, as only end-to-end testing holds them together.

The key aspect of a choreography is the *delivery notation* (~> in Chorex), which describes a communication between two actors. For example, the term Alice.e ~> Bob.v means that the actor Alice computes expression e and sends the result value to actor Bob to be stored in variable v.

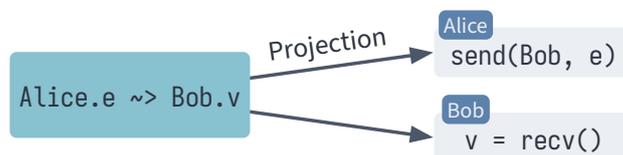

The expression e is *located* at actor Alice. Likewise, the variable v is located at Bob: Bob can use v in subsequent expressions, but Alice cannot. This is enforced by the compiler; it ensures that actors can only access information explicitly sent to them and that private computations remain local to that actor.

With choreographies, classes of communication errors become unrepresentable. Sends and receives cannot be mismatched because they are paired by design through delivery notation. Deadlocks cannot occur because lexical scope rules them out. For example, our language Chorex reports a compile-time error for the would-be deadlock below because the variable A.val is used before it is bound:

```
4  defchor [A, B, C] do # choreography for 3 actors     ERROR: undefined variable "val"
5      def run() do                                     |
6          A.val ~> B.val                               | A.val ~> B.val
7          B.val ~> C.val                               | ^^^
8          C.val ~> A.val                               |
9  end                                                  +- deadlock.exs:6 A.run/1
```

**(a)** Autocomplete in an actor lists functions required by the choreography.

**(b)** Missing knowledge-of-choice annotations lead to a static error during projection.

**Figure 1** Examples of language integration in Chorex.

Languages that support choreographic programming are on the rise and quickly growing to support full-featured programs. Choral [28] brings choreographies to Java and recently added interoperability with legacy code to enable an Internet Relay Chat (IRC) implementation [38]. MultiChor, for Haskell, introduces a dynamic approach to projection that has been ported to Rust and TypeScript [3]. These and other implementations (surveyed in Section 6) have taken great strides toward practical choreographic programming. However, choreographic languages fail to address all of the seminal *eight fallacies of distributed computing* [54, 63]: (1) the network is reliable, (2) latency is zero, (3) bandwidth is infinite, (4) the network is secure, (5) topology doesn't change, (6) there is one administrator, (7) transport cost is zero, and (8) the network is homogeneous. Languages meant to support distributed programming must be robust against all of these critical issues.

In this paper, we present a language—Chorex—that addresses a core aspect of fallacy (5), *topology doesn't change*, namely, *failing actors*. If an actor fails, Chorex can replace the actor and reset the choreography to a previous valid state. Chorex achieves this goal through a novel projection strategy and runtime monitoring. In a choreography, programmers write checkpoint/rescue blocks to specify recovery behavior. During projection, the compiler translates checkpoint blocks into code to make each actor checkpoint its state, as well as code to prevent actors from advancing into an un-recoverable position further in the choreography. At runtime, a supervisor restarts actors as needed, restores checkpoint state, and shares the address of the new participant via out-of-band messages (i.e., messages that are not specified in the choreography itself) to other actors.

In the following minimal example, Alice crashes due to division by zero. After restarting, a new Alice is able to exchange messages with Bob:

```
1 checkpoint do
2     Alice.f(1 / 0) ~> Bob.y
3 rescue
4     Alice.f(1) ~> Bob.y
5 end
6 Alice.(2 + 2) ~> Bob.sum
7 Bob.(sum + sum) ~> Alice.result
8 Alice.result
```

Chorex brings choreographic programming to Elixir. The implementation is notable because it follows the *languages as libraries* [52] design method to achieve a high level of integration with the standard Elixir toolchain. Figure 1 presents two benefits of language integration:

- Figure 1a shows that functions required by a choreography appear as suggestions when a programmer edits an actor module. The pictured suggestion appears in the context of an actor module named `MyAlice` that fills the role named `Alice` from a choreography named `Demo.Chorex` (not pictured). There are two required functions: `decrypt` and `priv_key`.
- Figure 1b shows a tooltip box with a compile-time error due to missing knowledge-of-choice annotations. This sort of error illustrated in Figure 1b is not detectable in choreographies implemented as runtime libraries because it requires two passes over the input [3, 46]. Either a bespoke compiler or Chorex-style metaprogramming is needed.

Both affordances are IDE-agnostic because they leverage the Elixir language server. Our IDE of choice happens to be Emacs, but Neovim or VSCode users would see similar tooltips.

Concretely, the Chorex compiler is an Elixir macro that analyzes source code, projects the code to actor implementations, and propagates source locations to enable informative error messages. Additional features of Chorex include first-class functions and out-of-order message receives.

**Outline** This paper begins by describing Chorex (Section 2), with emphasis on its novel support for restartable actors, and follows with a close look at the Chorex implementation and how it achieves language integration through metaprogramming (Section 3). Next is a series of examples of Chorex in action, including a TCP socket server and an implementation of the Secure Remote Password protocol (Section 4), and a performance evaluation of the overhead induced by the `checkpoint`/`rescue` recovery mechanism (Section 5). The paper concludes with a survey of the rapidly-evolving area of choreographic programming (Section 6) and a brief discussion (Section 7).

**Notation** For readability and to save space, code listings in this paper make two abuses of Elixir notation. First, they often omit `end` delimiters, which are required to close blocks opened by `def ... do` and other forms. Second, they omit parentheses around located expressions, writing `A.e` rather than the preferred `A.(e)`, which cooperates better with the Elixir autoformatter. Refer to the artifact for runnable Elixir code [62].

## 2 Elements of Chorex

Chorex is a domain-specific language for choreographic programming in Elixir [48, 50]. Elixir is the chosen target language for several reasons. First, it compiles to the Beam VM, the Erlang virtual machine, and thus has access to primitives that support low-latency, distributed, fault-tolerant systems. These primitives have enabled fast prototyping of choreographic features. Second, Elixir has a large userbase to engage with in future work. Third, Elixir comes with a hygienic macro system. Thanks to macros, Chorex is implemented as a library and integrates smoothly with the Elixir build system, Mix [21], and package manager, Hex [20].

Key design principles of Chorex include the following:

- Maintain a smooth Elixir workflow by implementing the choreography language and projection through metaprogramming.
- Manage actors using a standard Elixir supervision tree.
- Provide custom mailboxes and control stacks for actors to handle out-of-order messages, messages from the supervisor, and recovery.
- Use Elixir syntax and programming idioms to shape the "look and feel" of Chorex.

This section explains user-facing aspects of the language design, including how to write a choreography (Section 2.1), how to implement an actor (Section 2.2), and the framework for monitoring and supervision (Section 2.3).

### 2.1 The Choreography Language

To write a Chorex choreography, define a module, import Chorex, and use `defchor`:

```
1 defmodule SampleChor do
2     import Chorex
3     defchor [Alice, Bob, Carol] do
4         ...
5 end
```

The `defchor` macro expects two input forms: a list of actor roles (in CamelCase, to match the Elixir convention for module names) and a block of code. The block of code contains a sequence of function definitions (`def`). One function named `run` must be included; this function is the entry point to the choreography. Nothing other than function definitions are allowed. Each `def` projects to several variant functions, one for each actor. A `defchor` form outputs a standard Elixir module named `Chorex` that provides code and an API for actors. Expanded code thus has the following shape:

```
1 defmodule SampleChor do
2     import Chorex
3     defmodule Chorex do
4         ...
5 end
```

Other modules must refer to this macro-defined module (`SampleChor.Chorex`), either to implement an actor or to start the choreography.

**Located Expressions and Values** Variables and expressions in a choreography must be *located* at an actor using the syntax `Actor.e` (used in this paper) or `Actor.(e)` (preferred in practice, because it cooperates with Elixir's `mix format` tool). An actor can access variables located only on that actor; other actors' located variables are kept separate. When an expression is located on actor e.g. `Alice`, that expression will be computed on the process for the actor `Alice` when the choreography runs.

Arguments to standard functions must be located as well. The header below expects a field named arg at the actor `Alice` and another field, also named arg, at the actor `Bob`:

```
1  def some_function(Alice.arg, Bob.arg) do ...
```

Projecting this choreographic function results in two standard Elixir functions: one in a module for `Alice`, and the other in a module for `Bob`. Each function expects two arguments, but `Alice`'s projection will ignore the second argument, and `Bob`'s projection will ignore the first. At runtime, when the function is called, calls to `some_function` on node `Alice` will pass a dummy value as the second value, and similarly for `Bob`.

There is one exception to the rule that every value must be located: *function arguments* to a higher-order function are not located. Such functions have a representation on every actor in the choreography.

**Delivery: Sending and Receiving Messages**   Delivery notation (send ~> recv) sends a value from one actor to another. In Chorex, the sender can prepare any located expression: variables, function calls, arithmetic, and other expression forms are valid on the left side of a send. The receiver can use Elixir pattern matching to bind variables:

```
1  Alice.{:answer, 42} ~> Bob.{:answer, the_answer}
```

**Conditionals and Knowledge of Choice**   Chorex repurposes `if` expressions from Elixir for choreographic conditionals (with one change: Chorex requires an `else` branch; multi-way conditionals are future work). In the following example, `Alice` is the *deciding actor* for this conditional, as the branch hinges on the result computed at `Alice`. The projection for `Bob` inserts a receive to wait for a *knowledge of choice* message to know which branch to take:

```
1  if Alice.make_decision() do
2      Alice.yes_branch() ~> Bob.d1
3      Bob.report(d1)
4  else
5      Alice.no_branch() ~> Bob.d2
6      Bob.report(d2)
7  end
```

An `if` in Chorex need not appear in tail position (unlike, e.g., Pirouette [31]).

Chorex does not (yet) infer the actors in a conditional, and thus by default shares knowledge of choice with *every other actor* in the choreography. To limit the notified actors, a programmer can add a `notify:` annotation:

```
1  if Alice.make_decision(),
2      notify: [Bob, Carol] do ...
```

When a `notify:` fails to include all necessary actors, Chorex raises a compile-time error as it projects code for each actor. Below, `Carol` is missing a notify:

```
1  defchor [Alice, Bob, Carol] do
2      def run(Alice.msg) do
3          if Alice.decrypt(msg, priv_key()), notify: [Bob] do
4              Bob.notify_success()
```

```
5              Carol.foiled()
6         else
7              Bob.notify_failure()
8              Carol.success()
```

The error output explains the problem with an accurate line number:

```
== Compilation error in file bad_branch.ex ==
** (CompileError) bad_branch.ex:3: Branches differ for actor Elixir.Carol; `if' block needs to notify
```

**Error Rescue and Restarts** Chorex adapts Elixir's checkpoint/rescue blocks to handle errors that may arise in actor code. (Unlike in Elixir, rescue declares a block and not a match clause.) If Alice or Bob were to fail in the following checkpoint block, both actors would execute the rescue block with the failing actor restored as a new actor instance with recovered state:

```
1 checkpoint do
2     Alice.dangerous_operation() ~> Bob.x
3     Bob.success(x)
4 rescue
5     Alice.safe_operation() ~> Bob.x
6     Bob.fallback(x)
7 end
```

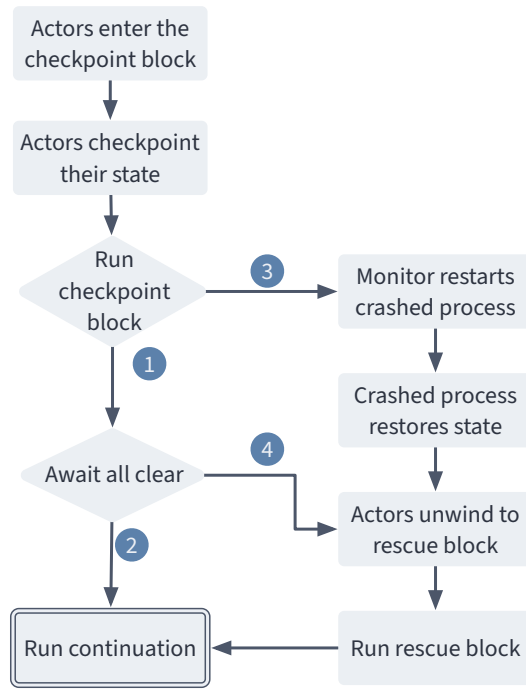

**Figure 2** Chorex checkpoint/rescue logic.

Figure 2 presents a flowchart description of checkpoint/rescue semantics. Every actor works through this flowchart by communicating with a runtime monitor, which watches for crashes. At the start of the checkpoint block, each actor checkpoints its state. This checkpoint is used to enter the rescue block, if needed. If all goes well for an actor (1), it pauses at the end of the checkpoint/rescue in case another actor crashes. If all actors pass (2), the monitor sends an "all clear" signal indicating that it is safe to proceed into the continuation of the block. If an actor crashes, (3) the monitor restarts the failed actor and notifies all actors to reset their state (4) and enter the rescue block. In either case, when the checkpoint/rescue block is done, the monitor drops the actors' checkpoints for this block and the actors proceed through the rest of the choreography.

**Actor-Local Variables** An actor can bind variables using Elixir's with notation. For example, here Alice creates a located variable x:

```
1 with Alice.x <- compute_value() do …
```

**Run Results**   When actors finish executing the main `run` function, they each send a value to the mailbox of the calling process. This value defaults to `nil`.

### 2.2  Actor Interface

Chorex projects a choreography into actor modules that handle communication, but not actor-specific functionality. Actor implementation modules must define all the local functions that a choreography needs. For example, in the choreography code below, the first line calls a function `get_money()` located on the Alice actor, the second line asks Bob to `fetch_apples`, and the third line asks Alice to `fetch_sugar` and `bake_pie`:

```
1  Alice.get_money() ~> Bob.payment
2  Bob.fetch_apples(payment) ~> Alice.apples
3  Alice.bake_pie(apples, fetch_sugar())
```

An implementation for `Alice` must provide each function in the wishlist:

```
1  defmodule AliceImpl do
2      use SampleChor.Chorex, Alice
3
4      def get_money(), do: …
5      def fetch_sugar(), do: …
6      def bake_pie(apples, sugar), do: …
7  end
```

To guide actor implementation, Chorex gathers the set of local functions for each actor during projection and creates an interface specification—called a *behaviour* in Elixir parlance—that an implementing module must contain. Elixir issues compile-time errors if an actor implementation does not satisfy the behaviour. Implementing modules must have `use SampleChor.Chorex, ActorRole`, with `ActorRole` replaced with the name of the actor to implement. This `use` declaration expands into a behaviour declaration for that actor, as well as some utility module imports needed by Chorex. Aside from this, there is nothing Chorex-specific in an actor implementation module.

**Language Server Integration**   Chorex provides guidance to implementation modules in the form of language server tooltips, illustrated in Figure 1a. These are enabled through metaprogramming and cooperation with Elixir's behaviour mechanism. In particular, the line `use SampleChor.Chorex, Alice` expands at compile time to code that glues this implementation module to the choreography's projected module for Alice.

**Starting a Choreography**   With a choreography defined in the module `SampleChor` as well as implementations for each of the actors, all that is left is to instantiate the choreography. The Chorex library defines a `start` function which takes the module name of the choreography, a map associating each actor to an implementation module, and a list of arguments to pass to the choreography entry point, i.e. the `run` function.

```
1  Chorex.start(SampleChor.Chorex,
2               %{Alice => AliceImpl, Bob => BobImpl, Carol => CarolImpl},
3               ["Hello␣from␣Alice"])
```

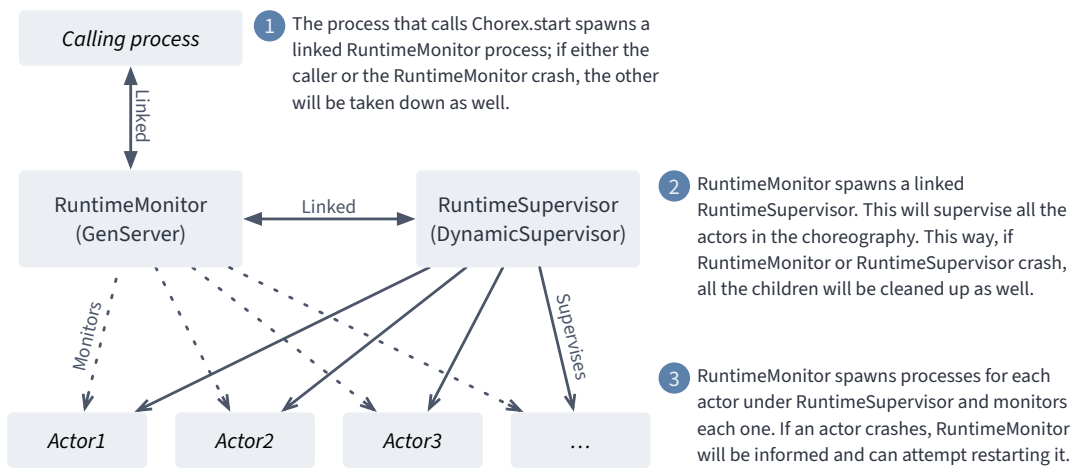

**Figure 3** Chorex creates one monitor and one supervisor for each choreography.

Chorex.start starts all the actor processes, broadcasts the network configuration so actors can find each other, and sets up the supervision tree.

## 2.3 Supervision Protocol

Chorex leverages Elixir/Erlang process monitoring to supervise choreographic actors. In Elixir, when process A monitors another process B and B exits, process A receives a message that describes how B terminated (e.g., normal exit vs. crash). Chorex creates monitoring links to build a supervision tree in the style of Figure 3 every time a program starts a choreography.

The supervision tree adds two processes to a choreography, in addition to the calling process and actor processes: a RuntimeMonitor and a RuntimeSupervisor. The Monitor, Supervisor, and calling process are linked together such that if any one process crashes, the entire choreography terminates. The Supervisor's job is to perform this cleanup on the actors. The Monitor's job is to watch for actors that crash, restart them, and send updates to reconfigure the network.

Technically, the Supervisor implements an Elixir behaviour called DynamicSupervisor [11]. The Monitor implements the GenServer behaviour [12]. Each actor is a GenServer as well; Section 3 explains the significance of GenServers.

**Restarting a Crashed Process** When an actor enters a checkpoint/rescue block, it creates a checkpoint that includes the control stack, message inbox, and bindings for local variables. The actor sends the control stack and variable bindings to the Monitor process. If a crash occurs, the Monitor spawns a new process, which has no state initially, and restores the missing pieces. After reinstating an actor process, the Monitor alerts all other actors that a crash has occurred.

Checkpoints are stored in the Monitor for convenience. They could easily move to the filesystem, a database, or even another process.

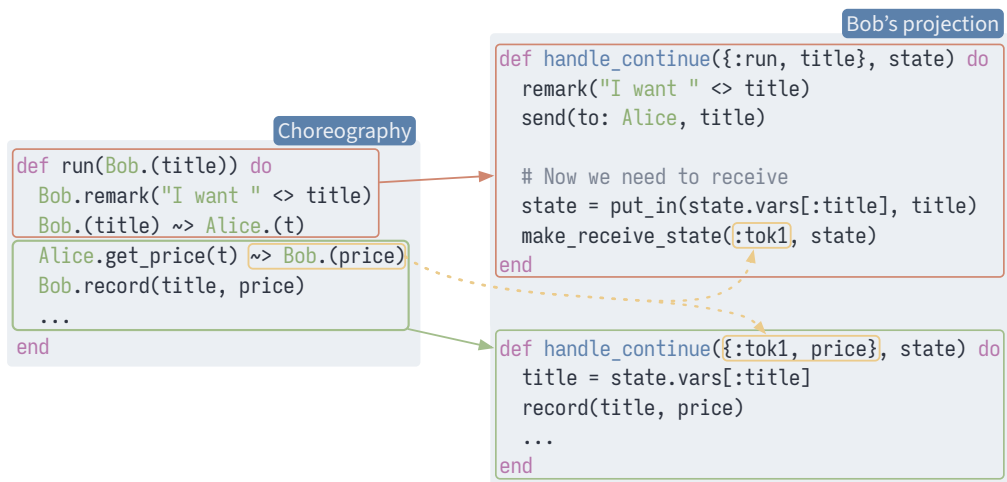

**Figure 4** Projection splits actor code into a set of callbacks: one per receive.

**Handling Out-of-Band Recovery Messages** Chorex manages its own call stacks and inboxes instead of relying on Elixir's mechanisms to enable out-of-band message receives. An out-of-band message is a message that does not correspond to any ~> in the choreography. When an actor crashes, the monitor broadcasts a message that the most recent checkpoint block failed along with a token indicating which stack frame the actors must unwind to. Actors handle this message by unwinding their control stacks and resetting their environments to the frame that transfers control to the rescue block. These reset messages are out-of-band because they bypass the queue that inter-process messages must normally go through.

**Can a Restarted Process Miss Messages?** Any messages that arrive while an actor is executing a checkpoint block go to its inbox, but not to its checkpointed state, and will be lost if the actor crashes. However, this is not problematic. The messages that arrive in a checkpoint block must have originated from other actors *within the same checkpoint block* because sends and receives are matched up and and all actors must complete the checkpoint before any one can continue execution. Forcing actors to wait at the end of a checkpoint also ensures that they are available to unwind if a rescue is needed; without synchronization, an actor might terminate early.

## 3 Implementation Highlights

Chorex's high-level strategy for enabling restarts and language integration is applicable to other choreographic languages, despite our focus on Elixir. This section discusses key aspects of the implementation to facilitate adaptation.

Chorex translates choreographies to sets *of sets* of message-passing functions (Figure 4). Each function in the choreography projects into several variants, one for each actor, and each variant is in turn split into several functions (distinguished by generated tokens) corresponding to receive-separated chunks of the choreography. Getting these functions to cooperate required a number of significant components: a tailored message format (Section 3.1), a method of organizing component functions (Section 3.2), and an implementation of the compiler as a metaprogram to maximize host-language integration (Section 3.3). We conclude this section with lessons learned from the effort (Section 3.4).

### 3.1 Message Format

There are two categories of messages that a Chorex actor needs to handle: choreography messages and control messages. *Choreography messages* go between actors, and originate from the sends and receives in a choreography. *Control messages* come from the Chorex runtime, and propagate knowledge of choice, notify actors of crashes and network updates, and synchronize actor execution of `checkpoint`/`rescue` blocks.

Most messages have a 3-tuple format: `MSG = {message_type, civ_token, payload}`

The `message_type` field may have one of five possible values, which determines both the payload and the tuple shape of the remaining message:

- `:chorex`, for a choreography message. The corresponding payload is a message (any Elixir value) created by one actor and intended for another actor.
- `:choice`, for knowledge-of-choice. Payload is a Boolean choice value.
- `:revive`, for error recovery. This type of message is a 2-tuple; it does not include a CIV token. Payload is the new actor state to install.
- `:recover`, for error recovery. This type of message is sent to actors that did not crash, and asks them to unwind to the nearest `rescue` point and reset their environments. Payload represents the new network configuration.
- `:barrier`, for synchronization at `checkpoint` blocks. Indicates that all actors successfully completed the `checkpoint` block and may proceed into the continuation.

The `civ_token`, inspired by Ozone [41], preserves communication integrity in the presence of out-of-order messages. Each CIV token is a 4-tuple:

```
1  CIV = {session_token, metadata, sender, receiver}
```

The `session_token` is a UUID generated when the choreography is instantiated (during `Chorex.start`). Every instance of a choreography has its own session token. The `metadata` field describes the source-code location within the choreography where this message originated (i.e., the original `send ~> recv`). Crucially, both parties involved in the message receive equal `metadata` values, and different `send` sites lead to different `metadata` values. The `sender` and `receiver` components are the role names of the actors involved. All together, a CIV token ensures that a message goes only to the intended destination.

### 3.2 Actors as Server Processes

Chorex projects actors to Elixir GenServer behaviors [12, 23] rather than straight-line processes. This choice solves the following problems with default Elixir processes:

1. no way to prioritize messages from the Monitor process (Figure 3),

■ **Listing 1** Example GenServer that implements a counter

```
1  defmodule Counter do
2    use GenServer
3    def init(start_count), do: {:ok, start_count} # final element is server state
4    def handle_cast(:increment, count), do: {:noreply, count + 1}
5    def handle_call(:get_count, _, count), do: {:reply, count, count} # 2nd element goes to caller
```

2. no way to supervise for crashes.

GenServers—in contrast to straight-line processes—can be supervised and monitored. Additionally, they can handle messages that arrive in any order and can give priority to messages from the Monitor process. A straight-line process would have to anticipate every possible out-of-band message (e.g. those arriving from the Monitor process) at *every* receive. GenServers have their own drawbacks, e.g., complicated variable scope, but Chorex works around these problems.

**GenServer Primer**   GenServer (short for "Generic Server") is an Erlang library that makes it easy to build processes that manage state and can respond to ad-hoc messages. Elixir inherits GenServers from Erlang. To behave as a GenServer, a module must define an init function that returns a value representing the server state, and callbacks to handle incoming messages. There are three kinds of callbacks that deal with messages: handle_call, handle_cast, and handle_info. These callbacks can expect at least a message and state value as input, and must return a new state value. A handle_call receives a process ID as well, and must include a reply to this process in its return value.

As an example, the GenServer in Listing 1 implements a counter. The initial state is the number start_count. One callback, handle_cast, awaits messages with the tag :increment and a numeric value; it adds the input value to the counter state. Another callback, handle_call, awaits :get_count messages. It logs information about the request and returns a 3-tuple with two copies of the state. One copy is a reply to the sender process and the other copy is the new state (same as before the call). To start an instance of this module with an initial count of zero, call GenServer.start(Counter, 0).

Multiple processes can interact with a GenServer at the same time. The GenServer handles each incoming message, one at a time. This behavior enforces linearity so that concurrent processes can interact with shared state in a coherent way.

**Challenges**   The key challenge of GenServers as a target for projection is that, in order to handle out-of-band messages, every message that an actor can receive must be anticipated with a callback. This means that receives must be split across several functions as shown in Figure 4. Consequently, variables defined earlier in a choreography must become part of the GenServer state in order to reach later parts of the choreography. A second consequence is that actors can no longer use the Elixir call stack to handle function calls. Chorex manages this with its own actor-specific control stacks in the GenServer state. These stacks also allow the GenServer to alter its control flow in response to out-of-band messages.

**Correct Scope via Live Variable Analysis**   Below, two actors send a message to `Bob`:

```
1  Alice.one() ~> Bob.x
2  Carol.two() ~> Bob.y
3  Bob.(x + y)
```

Projection for the `Bob` actor introduces two callbacks, one for each receive. A direct but incorrect projection would use the variables `x` and `y` directly. This is wrong because `x` is not in scope for the second callback, which needs to return a sum:

```
1  # WRONG projection for Bob
2  def handle_info({Alice, x}, state), do: ...
3  def handle_info({Carol, y}, state), do: x + y # WRONG
```

Chorex builds a correct projection by tracking the set of live variables through a choreography. At runtime, Chorex stores a map from variable to values in GenServer state, and reads from the map as needed:

```
1  # CORRECT projection for Bob
2  def handle_info({Alice, x}, state) do
3      state = put_in(state.vars[:x], x)
4      ...
5
6  def handle_info({Carol, y}, state) do
7      x = state.vars[:x]
8      x + y
```

Determining free variables in Elixir has some subtleties. For example, the match expression `[x, ^y, x] = make_list()` contains three variables, `x`, `y`, and `x` again, but binds only one (`x`). All variables with the same name in a pattern must bind to the same value, so the second `x` does not introduce a second binding. The variable `y` is *pinned* with the `^` prefix, meaning that the *value* of `y` should be matched in the pattern. If, for example, `y` were bound to `42`, this pattern would match lists like `[1, 42, 1]` or `["hi", 42, "hi"]`. Chorex reuses Elixir's tree-walking API to facilitate analysis, but it implements a custom set of rules to find free variables.

**Receives and Function Calls**   The second challenge of projection to GenServers is that function calls in a choreography no longer map cleanly to function calls in the Elixir output. An actor might call a function that receives several messages. Each receive will introduce a new callback. In the example below, the function `test_system` receives one message for each of the actors `Mike` and `Joe`:

```
1  def run() do
2      Joe.(:begin) ~> Mike.start_message
3      with Joe.response <- test_system() do
4          Joe.(String.length(response))
5
6  def test_system() do
7      Joe.("Hello Mike") ~> Mike.("Hello " <> my_name)
8      Mike.("Hello Joe, you said #{my_name}") ~> Joe.reply
9      Joe.("Received " <> reply)
```

To recover typical call/return behavior, each actor tracks a stack of call frames in the GenServer state. When a callback finishes, the GenServer inspects the top stack frame to decide where to go next. It then invokes the next callback which corresponds to a return in the source language.

When projecting the `run` function above, Chorex does not know whether the call to `test_system` will perform any receives. (Higher-order functions make it impractical to statically track which functions do receive.) At each function call, Chorex thus creates a unique token to identify the call site and its continuation. This token is used in two places: first, it goes onto the control stack in the GenServer state; second, it appears in the argument specification of the callback that holds the continuation code.

**Actor State** Each Chorex actor keeps the following state, which gets passed between every message handler in the GenServer implementation:

- a queue of choreography messages to be processed,
- a stack of control frames (to recover receives and function calls),
- a map of live variables,
- the `session_token`,
- a map representing the network configuration, and
- a reference to the implementing module.

Chorex actors also share some functionality via a runtime module. The runtime can: push new messages on a queue as they arrive, inspect the control stack to determine which message in the inbox is needed next, and unwind execution stacks.

**Minimizing Memory in Checkpoints** Actor stacks grow linearly as they descend into recursive function calls. If an actor checkpoints state at each level, the set of saved states would grow quadratically as depth increases; this happens when a recursive call is inside the `checkpoint` block, as seen in the `Nest-10k` benchmark in Section 5. To prevent excessive memory usage, the Monitor process saves deltas of each actor's stack. This makes checkpointing and restoring state slightly more expensive, but this cost is offset by reduced memory usage.

### 3.3 Projection via Macro Expansion

Chorex uses Elixir's macro system to embed a choreography language. Macros expand during compilation, which allows Chorex to perform static checks such as sufficient knowledge-of-choice propagation. Macros are also part of the standard Elixir toolchain, which makes for a seamless workflow. *No extra build steps are needed.*

With its macro implementation, Chorex reuses many affordances of the host language. Local expressions get lowered as-is, making the entirety of Elixir available. Macro hygiene means Chorex users do not have to worry about macro implementation details leaking out, and Chorex itself is, in principle, macro extensible.

**Defchor Internals** The `defchor` form is a macro that takes a list of actor roles and a block of choreography code. It projects the choreography body for each of the actors.

Projection takes an actor role and a sequence of expressions and returns three values: (1) a sequence of expressions, representing the actor's view of the expressions; (2) a list of function clauses, which `defchor` will splice into the GenServer for the actor; and (3) a list of function specifications, which will be required of actor implementations. With these pieces, the `defchor` macro generates a module for each actor that contains code to realize that actor's communications as well as a behaviour spec which actor implementations (Section 2.2) must satisfy.

Elixir AST nodes include metadata about source code, including line and column numbers. Chorex uses this metadata whenever possible in expanded code to ensure that error messages get reported in terms of the source language. For example, for the following faulty code: `Alice.one(bad_variable_name) ~> Bob.x` macro output causes the Elixir compiler to report a readable error:

```
error: undefined variable "bad_variable_name"
    Alice.one(bad_variable_name) ~> Bob.x
```

### 3.4 Reflections

GenServers as a compilation target enabled flexible, out-of-order and out-of-band message receives. This critical ability was well worth the pain of having to implement custom mailboxes, live variable analysis, and execution stacks.

One major issue in Elixir's macro system is its lack of support for pattern matching on quoted syntax. Although Elixir has excellent pattern matching for values [22], matching on the raw AST terms becomes verbose and cumbersome for non-trivial matches. A tool similar to Racket's `syntax-parse` [17] would make macros easier to write.

Chorex projects choreographies into one module and requires actor implementations to provide application-specific details in a separate module. This design allows the reuse of one protocol across several implementations, and gives Chorex a natural way to reuse Elixir language tooltips. Implementation modules get behavior injected into them via the call `use ChorModule.Chorex, ActorName` that transparently turns them into GenServer modules. This approach works well in our experience; in the future, however, moving to a purely behaviour-based system for local implementation might make things more in line with some Elixir conventions.

## 4 Chorex in Action

This section demonstrates the use of Chorex on motivating examples. The main examples are secure remote password authentication (Section 4.1) and a TCP socket server (Section 4.3). We also include a zero-knowledge protocol for a discrete logarithm (Section 4.2) and, of course, a bookseller example (Section 4.4).

### 4.1 Secure Remote Password

Secure Remote Password (SRP) [64] is an authentication method based on zero-knowledge-proofs [29]. Our Chorex implementation drives a simple command-line application that lets one user register a password and then serves login requests:

```
iex(1)> ZkpLogin.register_srp()
    [New User SRP] username: alice
    [New User SRP] password: jabberwocky
    Server responds {:registered, "alice"}
    Client responds :registered
    :ok

iex(2)> ZkpLogin.login_srp()
    [Login] username: alice
    [Login] password: dormouse
    Server responds {:fail, :reject_client_digest}
    Client responds {:fail, :server_rejected_digest}
```

The choreography has two run functions corresponding to the registration and login phases. The login function expects no input at either actor. The registration function expects a username and password located at the client, and a :register token located at the server—merely to differentiate the projected version of this server function from the server's login function. In general, Elixir encourages the use of overloaded functions distinguished by arity and argument patterns. Chorex supports overloaded functions as well, but with the subtle requirement that the functions for each actor must have distinct argument patterns *after projection*.

The code below shows the registration run function; Figure 5 describes the logic of the login run function:

```
 1 defmodule Zkp.SrpChor do
 2     import Chorex
 3
 4     defchor [SrpServer, SrpClient] do
 5         def run(SrpClient.{uname, pwd}, SrpServer.(:register)) do # register
 6             SrpServer.get_params() ~> SrpClient.{salt, g, n}
 7             with SrpClient.v <- SrpClient.gen_token(uname,pwd,salt,g,n) do
 8                 SrpClient.{uname, salt, v} ~> SrpServer.{uname, salt, v}
 9                 if SrpServer.register(uname, salt, v) do
10                     SrpServer.{:registered, uname}
11                     SrpClient.(:registered)
12                 else
13                     SrpServer.({:error, :no_registration, uname})
14                     SrpClient.({:error, :no_registration})
15
16         def run() do … # login
17 end
```

An important property of SRP is that secret information, such as the password on the client or the randomly-generated session secret $b$ on the server, never get transmitted. By contrast to a traditional distributed implementation where client and server code are separate, such as srp-elixir [45], this property is easier to verify with

1. Client sends user_id to server.
2. Server finds user's salt $s$ and auth key $v$, uses constants $g$, $N$ to compute $k = hash(g,N)$, generates random secret $b$, computes $B = kv + g^b$, and sends $g$, $N$, $s$, $B$ to client.
3. Client generates random secret $a$ and computes several values, including a session key:
   $A = g^a$
   $k = hash(g,N)$
   $u = hash(A,B)$
   $K = (B - kg^x)^{a+ux}$
   $M_1 = hash(A,B,K)$
   Client sends $A$, $M_1$ to server.
4. Server computes $K = (Av^u)^b$ and checks that $hash(A,B,K) = M_1$. Server sends $M_2 = hash(A, M_1, K)$ to confirm session key.

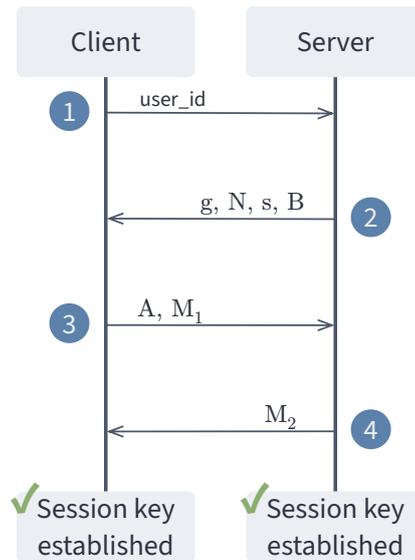

**Figure 5** Secure Remote Password login protocol to establish session key $K$.

a choreography. The choreography shows what values cross between actors. These values may depend on local functions, such as gen_token, whose correctness must be established separately; nevertheless, the choreography narrows the room for error.

During the login phase of SRP, there are several rounds of communication that take place between server and client to establish a session key $K$. The choreographic function that implements these exchanges is roughly twice as long as the registration function but uses similar language features (with, ~>, if), so we defer it to the artifact. It is enough to say that comparing the choreography against a high-level algorithm description, shown in Figure 5, is straightforward. For example, the login choreography has exactly four communication terms (~>), matching the figure.

### 4.2 Discrete Logarithm

For our second example, we have implemented a zero-knowledge proof protocol to convince a verifier that the prover knows the logarithm of a number in a finite field. This is similar to the SPR login (Section 4.1) but it (1) requires multiple challenge rounds for the proof, and (2) does not produce a random session key like the SRP protocol. Below is part of the authentication choreography, abbreviated for space. The full choreography is approximately 70 lines long and is included in the artifact.

```
1  defchor [Prover, Verifier] do
2    def run(Verifier.rounds) do
3      Prover.get_username() ~> Verifier.ident
4      # Prover looks up authentication parameters; y is validation key
5      with Verifier.creds <- Verifier.lookup(ident) do
6        if Verifier.creds do
7          with Verifier.{y, p, g} <- Verifier.creds do
8            Verifier.{p, g} ~> Prover.{p, g}
```

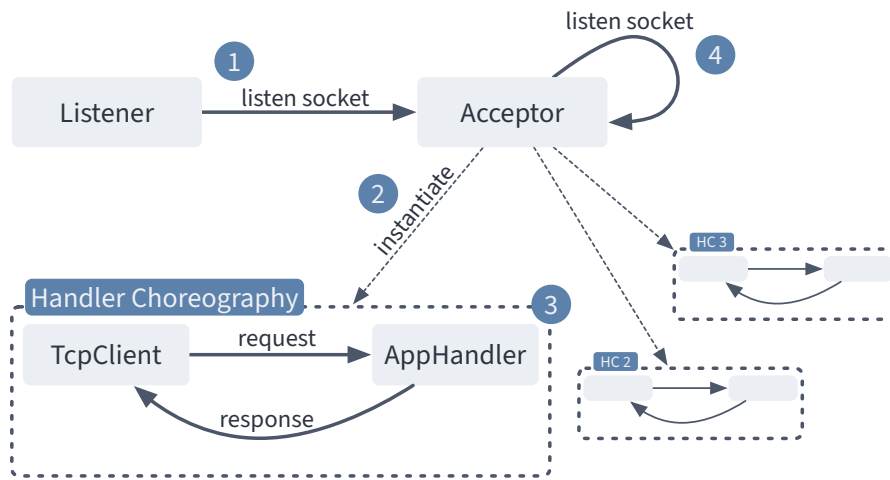

**Figure 6** Socket server architecture: (1) Listener sends a socket to Acceptor; (2) Acceptor waits for a TCPClient, then spawns a Handler Choreography; (3) TCPClient and AppHandler exchange messages; (4) Acceptor listens for a next client.

```
 9            round_loop(Verifier.{p, g, y}, Verifier.rounds, Prover.{p, g, get_secret(get_username())})
10       else
11         Verifier.(:bad_username)
12   # y is the validation token, x is the client secret
13   def round_loop(Verifier.{p, g, y}, Verifier.rounds, Prover.{p, g, x}) do
14     if Verifier.(rounds <= 0) do
15       Verifier.(:accept)
16     else
17       with Verifier.good_proof? <- do_challenge(Verifier.{p, g, y}, Prover.{p, g, x}) do
18         if Verifier.good_proof? do
19           round_loop(Verifier.{p, g, y}, Verifier.rounds - 1, Prover.{p, g, x})
20         else
21           Verifier.(:reject)
22   def do_challenge(Verifier.{p, g, y}, Prover.{p, g, x}) do
23     ...
```

The choreography uses two closely-knit helper functions: one that loops through several rounds of challenge and verification, and another that handles the logic of a single round. Compiling these functions to a core language that supports restarts was a key milestone in the development of Chorex.

### 4.3 TCP Socket Server

Our third example is a minimal socket server inspired by Thousand Island [51], a full-fledged socket server written in Elixir. There are two choreographies involved. A first choreography, between a Listener actor and an Acceptor actor, initializes a server that clients can connect to. A second choreography specifies interactions between a Client actor and an AppHandler actor. Figure 6 maps out the overall design. Crucially, the top-level choreography is able to start multiple instances of the inner Handler

Choreography. Every client that connects to the server gets forwarded to a unique choreography with an AppHandler actor.

**Top-Level, Listener Choreography** The listener choreography for our TCP server expects configuration data as input and immediately calls a helper function (located on the Listener actor) to acquire a socket connection. The Listener sends this socket to an Acceptor actor. At this point, the Listener's work is done. The choreography then enters a loop in which the Acceptor awaits and manages incoming connections.

```
1 defmodule Tcp.ListenerChor do
2     import Chorex
3
4     defchor [Listener, Acceptor] do
5         def run(Listener.config) do
6             Listener.get_listener_socket(config) ~> Acceptor.{:ok, socket}
7             loop(Acceptor.socket)
8
9         def loop(Acceptor.listen_socket) do
10            Acceptor.accept_and_handle_connection(listen_socket)
11            loop(Acceptor.listen_socket)
12 end
```

The following module implements the Acceptor actor. It imports the choreography above and, on the same line, specifies the actor role that it plans to implement (Acceptor). Inside the helper function, this Acceptor calls Chorex.start to invoke a choreography named Tcp.HandlerChor. The other arguments to Chorex.start are a map from actor roles to module names and a list of arguments to the HandlerChor's run function. In this case, the only argument is the socket:

```
1 defmodule Tcp.AcceptorImpl do
2     use Tcp.ListenerChor.Chorex, Acceptor
3
4     @impl true
5     def accept_and_handle_connection(listen_socket) do
6         {:ok, socket} = :gen_tcp.accept(listen_socket)
7         Chorex.start(
8             Tcp.HandlerChor.Chorex,
9             %{AppHandler => Tcp.HandlerImpl, TcpClient => Tcp.ClientImpl},
10            [socket]
11        )
12 end
```

If anything goes wrong at runtime, the Acceptor will exit cleanly and bring the Listener down as well. This exit behavior comes out of the box with Chorex.

**Inner, Handler Choreography** The second choreography describes an interactive loop. First, the AppHandler actor initializes a dictionary to track the total bytes sent by the client. Inside the loop, the Client sends a message to the AppHandler and the AppHandler updates its state, decides whether to continue, and sends a reply:

```
1 defmodule Tcp.HandlerChor do
2     import Chorex
```

```
 3
 4    defchor [AppHandler, TcpClient] do
 5        def run(TcpClient.sock) do
 6            loop(AppHandler.(%{byte_count: 0}), TcpClient.sock)
 7
 8        def loop(AppHandler.state, TcpClient.sock) do
 9            TcpClient.read(sock) ~> AppHandler.msg
10            with AppHandler.{resp, st2} <- AppHandler.run(msg, state) do
11                AppHandler.fmt_reply(resp) ~> TcpClient.resp
12                TcpClient.send_over_socket(sock, resp)
13                if AppHandler.continue?(resp, st2) do
14                    loop(AppHandler.st2, TcpClient.sock)
15                else
16                    TcpClient.shutdown(sock)
17                    AppHandler.ack_shutdown()
18 end
```

This choreography uses a Chorex if expression in which the AppHandler makes a choice that determines the future of the conversation. The TcpClient receives a *knowledge of choice* message from Chorex to know which branch to take in its own projected code.

### 4.4 More Examples: Higher-Order and Out-of-Order

Chorex has several other features inspired by prior work on choreographies. For one, it supports *first-class choreographic functions*. The program in Listing 2, adapted from the Pirouette paper [31], has a run function that executes either a one-buyer or two-buyer bookseller scenario by passing one function to another function. Both scenarios are inspired by the session types literature [6, 32].

References to functions, namely @two_party/1 and @one_party/1, are prefixed with an @ sign so that Chorex can increment the arity to account for an implicit argument representing the choreography state (Section 3.2), during compilation. This prefix is a slight twist on Elixir's standard &-sign prefix for function references. The suffix /1 is standard for Elixir; it describes the arity of the function.

A second important feature is *out-of-order message receives*. In the following example, the two messages sent to MainServer arrive when they are ready. The first send does not block and the second, being data-independent, can run immediately:

```
1 defchor [KeyServer, MainServer, ContentServer, Client] do
2     def run() do
3         ContentServer.getText() ~> MainServer.txt # may arrive 2nd
4         KeyServer.getKey() ~> MainServer.key # may arrive 1st
5         …
```

This feature is inspired by the Ozone language [41]. However, unlike the calculus that Ozone is based on ($O_3$), Chorex will not reorder two sends from the same actor, nor will it move expressions in or out of an if branch.

■ **Listing 2** Higher-order choreography that handles two classic bookseller scenarios.

```
1  defchor [Buyer, Contributor, Seller] do
2      def run(Buyer.include_contributions?) do
3          if Buyer.include_contributions? do
4              bookseller(@two_party/1)
5          else
6              bookseller(@one_party/1)
7
8      def bookseller(f) do
9          Buyer.get_book_title() ~> Seller.the_book
10         with Buyer.decision <- f.(Seller.get_price("book:" <> the_book)) do
11             if Buyer.decision do
12                 Buyer.get_address() ~> Seller.the_address
13                 Seller.get_delivery_date(the_book, the_address) ~> Buyer.d_date
14                 Buyer.d_date
15             else
16                 Buyer.nil
17
18     def one_party(Seller.the_price) do
19         Seller.the_price ~> Buyer.p
20         Buyer.(p < get_budget())
21
22     def two_party(Seller.the_price) do
23         Seller.the_price ~> Buyer.p
24         Seller.the_price ~> Contributor.p
25         Contributor.compute_contrib(p) ~> Buyer.contrib
26         Buyer.(p - contrib < get_budget())
27 end
```

## 5 Performance Overhead

Since actors checkpoint their state upon entering a `checkpoint` block and unwind their execution stack to enter a `rescue` block, it is important to benchmark the runtime overhead. We have tested with two realistic case studies of `checkpoint`/`rescue`, inspired by programs from Section 4, and several microbenchmark variants. Table 1 lists representative results. State Machine is based on the TCP choreography. Mini Blockchain computes hashes in a loop, similar to *challenges* in zero-knowledge protocols. Flat-10k is a microbenchmark of 10,000 iterations through a recursive function; in each iteration, two actors do some work in a single checkpoint block. Nest-1k and Nest-10k respectively run 1,000 and 10,000 iterations of a similar recursive function, but with a recursive calls inside the `checkpoint` and `rescue` blocks.

Each benchmark program comes in three versions: a plain version that does not use `checkpoint`/`rescue` and in which actors never crash; a non-crashing version (`chk`) in which actors must go through a checkpoint block, but still never crash (i.e., same control flow as plain version); and a crashing version (`chk + rescue`) in which actors do crash and the program must recover before it continues. In Table 1, the `chk` and `chk+rescue` columns report overhead relative to the plain, no-checkpoint version.

■ **Table 1** Overhead in two realistic programs and three microbenchmark configurations compared to counterparts with no checkpoint/rescue.

|  | chk | chk + rescue |  | chk | chk + rescue |
|---|---|---|---|---|---|
| State Machine | 1.01 × | 1.04 × | Flat-10k | 1.06 × | 5.44 × |
| Mini BlockChain | 6.76 × | 4.71 × | Nest-1k | 1.28 × | 1.95 × |
|  |  |  | Nest-10k | 3.48 × | 1.96 × |

All experiments ran on a single-user Apple M1 Pro with 32 GB RAM and 10 available cores, using Chorex 0.9.1, Elixir 1.18.0, and Erlang 27.2.

All benchmarks saw some overhead relative to baseline performance, as expected. In State Machine, Flat-10k, and Nest-1k, the crashing version of each benchmark imposes an equivalent-or-higher overhead relative to the baseline than the non-crashing version. This comes from the cost of starting up a new process, restoring its state, and broadcasting its new address to other actors in the system. The actors' states never grow very large, so the cost of checkpointing is low.

Both Mini BlockChain and Nest-10k feature deep call stacks stemming from recursive function calls *inside* the checkpoint/rescue blocks. This means that, in the non-crashing variants, the actors accumulate deep stacks in their process states which in turn increases the cost of checkpointing this state with the monitor process. In the crashing variants, the actor state does not get as deep because they slough off their checkpoint frames, the monitor does not accumulate states for every recursive call, and actors do not have to wait for an all-clear signal from the monitor to bubble out of the recursion.

Compile times scale linearly with the number of actors in the choreography. A choreography with 100 actors (already bigger than any practical example we were able to find) took approximately 11 seconds, and 1000 actors took approximately 2 minutes to compile. Elixir has a parallel compiler that Chorex does not currently use; employing this to speed compilation is future work.

## 6 Related Work

The design and implementation of choreographic languages has become a lively research area. Java, Haskell, Racket, Rust, and (now) Elixir all have third-party support for choreographies today, and most of these implementations appeared within the past year [3, 5, 9, 28, 37]. Choral is a notable exception with nearly a decade of engineering under its belt [28]. A first workshop on choreographies [1] and an introductory zine [2] appeared last year as well.

As Table 2 outlines, the overall expressiveness of choreographies as a protocol language is expanding. Functional (or higher-order [14, 31]) choreographies, which can use functions as first-class values, are standard. Restarts and network changes are unique to Chorex. Fully out-of-order execution (Full OoO) lets a choreography reorder code in any way that respects data dependencies. An efficient realization of (partial) reordering is in the Ozone API [41], which compiles to Choral. Census polymorphism allows abstraction over the number of participants in a choreography, analogous to variable-arity functions [19]. MultiChor, ChoRus, and ChoreoTS (which were introduced simultaneously [3]) are the first languages to support census polymorphism. Agreement types in Klor track the participants involved in a subroutine and thus provide a compositional way to infer knowledge-of-choice annotations. Chorex provides a modicum of reordering to avoid performance pathologies; messages sent to an actor arrive as soon as they are ready, instead of queuing. We have no plans at this time to implement full reordering. The other features in Table 2 are on the agenda for future work improving Chorex.

An orthogonal dimension is whether to implement choreographies as a standalone language or as a library. Libraries are simpler to implement and use, but limited in power. For example, HasChor [46] broadcasts knowledge-of-choice to all participants—turning every conditional into a choreography-wide sync point—because it cannot perform a two-pass static analysis (as in [31]). MultiChor and its relatives reduce the actors in each broadcast through a *conclave* mechanism. A thesis of Chorex, and of Choret [5], is that the best implementation of all comes through a meta-programmable host language such as Racket [18, 24]. Similarly, recent work improves HasChor with static projection via a Haskell compiler plugin [35].

Theoretical foundations of choreographies have a long history [7, 15, 16, 31, 41, 42]. The Pirouette calculus was the blueprint we followed to start Chorex [31].

Elixir is soon to acquire a full-featured gradual type system based on set-theoretic types [8]. Release v1.18 [53] infers types from match patterns and function bodies to report certain high-confidence warnings. However, users cannot write types and the type checker assumes the permissive dynamic type by default, limiting the guarantees that types provide. Extending this type system to gradually check choreographies is an exciting future direction.

Lastly, we mention areas that have close ties to choreographies. Multiparty session types [32, 33] (MPST) and secure multiparty computation (SMC) [43, 49] both coordinate distributed actors via a global protocol. Session types merely specify required behavior, giving developers the freedom to build a conforming implementation. One notable realization of session types is ElixirST, a type checker for a language of Elixir processes [26]. Writing programs that conform to session types can be challenging; techniques for API generation address the implementation burden [10], similar to how Chorex choreographies generate requirements for actor modules. SMC languages avoid the conformance question by generating code from a protocol. They are effectively choreographic languages, but designed for and constrained by security considerations. Restarts have been implemented and formally modeled in (at least) two session-types efforts: Session-CM [55, 56] and Links [25]. Session-CM (*cluster management*) is an alterate toolchain and runtime for Apache Spark, fully compatible with third-party Spark apps, that detects and replaces unresponsive nodes. Links extends binary session types with `try`/`catch` handlers. Chorex differs from these efforts through its close integration with a high-level source language (namely, Elixir), and its use of a compiler from choreographies to endpoint code. ScalaLoci is domain-specific language for fault-tolerant, distributed code [58]; it has a rich type system, including a notion of data placement, that future versions of Chorex may benefit from. Dezyne brings formal verification to concurrent industrial processes via a domain-specific

■ **Table 2** Recent advances in choreographic programming. Functions-as-values has seen wide adoption (Functional column). Other recently-proposed features have yet to permeate the landscape. Chorex adds error-restarts to the feature space.

|   | Functional | Restartable | Full OoO | Census | Poly | Agreement-$\tau$ |
|---|---|---|---|---|---|---|
| Choret [5] | ✓ | | | | | |
| HasChor [46] | ✓ | | | | | |
| ⋆ Chorex | ✓ | ✓ | | | | |
| Choral [13, 28, 41] | ✓ | | | ✓ | | |
| Klor [37] | ✓ | | | | | ✓ |
| MultiChor,ChoRus, ChoreoTS [3] | ✓ | | | ✓ | | |

language, simulator, and language server integration [4]. Verification in Chorex is an important next step. There is a long history of related work on verification for MPI programs to draw from as well [36, 47, 57].

Recent work on hybrid session types shows how to compose a global protocol from local, application specific protocols [27]. While Chorex allows choreographies to start other choreographies, and thus supports some composition, it cannot make static guarantees. Hybrid choreographies may be the way forward. Another closely-related work is the Corps calculus for hierarchical choreographies [30].

## 7 Conclusion

The essence of Chorex is a compiler from multiparty programs to stateless, message-passing processes. This compiler breaks new ground for choreographic programming with *restartable actors*, which are enabled through a checkpointing protocol and cooperative supervisor process, and with its *tight integration* to the host language, enabled by metaprogramming. It was not at all clear at the start of this project that an expressive choreographic language could be implemented as a meta-program, as opposed to a standalone compiler. The fact that Chorex works for Elixir indicates that other metaprogrammable languages, from Ruby [34] to Racket [18] to Rust [44], can follow suit by building a library for supervised processes (Section 2.3) and a pure-functional server runtime (Section 3.2). Extending the Chorex compiler protocol to support multiple languages in the same toolchain, with projection to actors written in different languages, would be a worthy challenge for the future.

**Data-Availability Statement**

The evaluated artifact for this paper contains all significant code examples, scripts for reproducing benchmark results, and a recent version of Chorex [61]. A newer version of the artifact with updated benchmarks and the latest version of Chorex is also available [62]. The latest version of Chorex is available on GitHub [60], and via the Elixir/Erlang package manager Hex [59].

**Acknowledgements** We thank Lee Barney and Dan Plyukhin for discussions that helped to frame this work, Ashton Snelgrove for introducing us to Deutsch's fallacies, and Andrew Hirsch for walking us through the details of Pirouette. We also thank the reviewers, whose feedback helped us improve our paper—in particular, their requests for a more detailed performance analysis helped us find and resolve inefficiencies around checkpointing actor states.

## About the authors

**Ashton Wiersdorf** (research@wiersdorfmail.net) is a PhD student at the University of Utah.
https://orcid.org/0000-0001-5524-7930

**Ben Greenman** (benjamin.l.greenman@gmail.com) is an assistant professor at the University of Utah.
https://orcid.org/0000-0001-7078-9287